\documentclass[twocolumn]{revtex4-1}
\usepackage{graphicx}
\usepackage{graphics}

\begin{document}

\title{Experimental ionization of atomic hydrogen with few-cycle pulses}


\author{M. G. Pullen$^{1,2,*}$, W. C. Wallace$^{1,2}$, D. E. Laban$^{1,2}$, A. J. Palmer$^{1,2}$, G. F. Hanne$^3$, A. N. Grum-Grzhimailo$^{4,5}$, B. Abeln$^4$, K. Bartschat$^4$, D. Weflen$^4$, I. Ivanov$^6$, A. Kheifets$^6$, H. M. Quiney$^7$, I. V. Litvinyuk$^2$, R. T. Sang$^{1,2}$, D. Kielpinski$^{1,2}$}

\address{$^1$ ARC Centre of Excellence for Coherent X-Ray Science, Griffith University, Nathan, QLD, 4111, Australia\\
$^2$ Australian Attosecond Science Facility and Centre for Quantum Dynamics, Griffith University, Nathan, QLD, 4111, Australia\\
$^3$ Atomic and Electronics Physics Group, Westf\"alische Wilhelms-Universit\"at, M\"unster, Germany\\
$^4$ Department of Physics and Astronomy, Drake University, Des Moines, Iowa, USA\\
$^5$ Institute of Nuclear Physics, Moscow State University, Moscow, Russia\\
$^6$ Research School of Physical Sciences, The Australian National University, Canberra, ACT, Australia\\
$^7$ ARC Centre of Excellence for Coherent X-Ray Science, University of Melbourne, Melbourne, VIC, Australia
$^*$Corresponding author: mgpullen@gmail.com
}

\begin{abstract}We present the first experimental data on strong-field ionization of atomic hydrogen by few-cycle laser pulses. We obtain quantitative agreement at the 10\% level between the data and an {\it ab initio} simulation over a wide range of laser intensities and electron energies.\end{abstract}

\maketitle

The interaction of intense few-cycle infrared laser pulses with matter induces tunneling ionization and subsequent quantum dynamics of freed electrons. Intense few-cycle pulses are difficult to generate and use because of the stringent requirements on dispersion control over a broad bandwidth. However, they offer unparallelled opportunities to reveal and control the electronic dynamics of atoms \cite{Baltuska-Krausz-as-atomic-control, Rudenko-Ullrich-correlated-multielectron-dynamics} and molecules \cite{Legare-Corkum-few-cycle-double-ionisation, Alnaser-Cocke-few-cycle-H2-control} and to generate isolated attosecond pulses in the extreme ultraviolet \cite{Hentschel-Krausz-as-metrology}. The few-cycle regime is particularly challenging for simulations, as intensities approaching $10^{15} \: \mbox{W}/\mbox{cm}^2$ can be reached before the ionisation response saturates. At these intensities, a photoelectron driven by intense long-wavelength radiation can travel a distance hundreds of times larger than the size of the parent atom and can have energies of many tens of eV, imposing stringent requirements on the simulation grid. {\it Ab initio} simulations in this regime can be carried out only for atomic H due to its simple electronic structure.

Here we describe an experiment on the interaction of intense few-cycle laser pulses with atomic hydrogen (H), the simplest of all atomic systems and the traditional test case for atomic physics. No data on H has previously been available in this regime of laser interaction. Previous strong-field experiments with atomic H \cite{Rottke-Welge-hydrogen-low-order-ATI, Paulus-Walther-hydrogen-ATI-expt} used relatively short-wavelength pulses that were many optical cycles in duration with maximum intensities of $10^{14} \: \mbox{W}/\mbox{cm}^2$. Our data show excellent quantitative agreement, at the 10\% level, with \emph{ab initio} simulation over a wide range of electron energies and laser intensities.

The experimental apparatus is composed of an atomic H beam interacting with a few-cycle strong-field laser (Fig. 1). The laser used is a commercial Femtolaser `Femtopower Compact Pro'. Each pulse has energy of~150 $\mu$J and the pulse repetition rate is 1 kHz. The spectral width of the laser is 150 nm at full width half maximum (FWHM) centered at 750 nm. The pulse duration at FWHM of the intensity envelope is 6.3 $\pm$ 0.2 fs at the interaction region, or alternatively $\sim 2.5$ optical cycles. An off-axis parabolic mirror of 750 mm focal length is used to focus the beam to a spot size of 47 $\mu$m $1/e^2$ radius. The laser carrier-envelope phase was not stabilized in these experiments.

The atomic H beam is created via collisional dissociation in a radio frequency (RF) discharge powered by a helical resonator \cite{Slevin-Stirling-rf-H-dissociator}. An RF signal at a frequency of 75 MHz and power of 8 W is applied to the resonator and the dissociation efficiency is determined via emission spectroscopy of the discharge. The atomic beam emerging from the discharge is $80 \pm 15$\% H atoms by number with the remainder being undissociated H$_2$. The atomic H beam passes through two apertures, producing a uniform-density beam of 0.5 mm diameter and $<2 \times 10^{-3}$ mrad divergence angle at the interaction region.

The electron detection system is composed of a series of electrostatic lenses that act to repel low energy electrons. Electrons passing the repeller are accelerated to $\sim 250$ eV before being detected by a channeltron. The channeltron is positioned in line with the laser polarization direction and is operated in electron counting mode.

\begin{figure}
\centerline{\includegraphics*[width=\columnwidth]{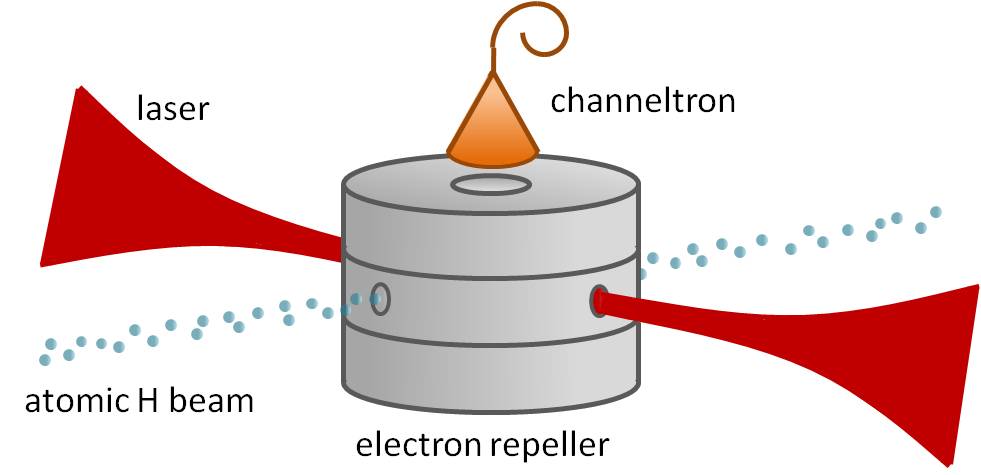}}
\caption{A beam of H atoms interacts with a few-cycle laser beam and photoelectrons are ejected with a wide range of energies. Electrons with energy above a cutoff value emerge from the repeller and are detected by a channeltron.}
\label{schematic}
\end{figure}

The experimental procedure involves measuring the electron yield over 10,000 laser pulses for a range of laser intensities and repeller voltages.  To obtain the electron yield for atomic H alone ($C_H$), we take electron counts for three different settings: with the dissociator on ($C_\mathrm{on}$), with the dissociator off ($C_\mathrm{off}$) and a background measurement taken with the atomic beam blocked ($C_\mathrm{bkg}$). The parameters are combined according to
\begin{equation}
C_H = (C_\mathrm{on} - C_\mathrm{bkg}) - (1 - \mu)(C_\mathrm{off} - C_\mathrm{bkg}), \label{hextract}
\end{equation}
where $\mu$ is the fraction of atomic H in the beam. The procedure is repeated for a range of peak intensities from 1.2 to 5.4 $\times 10^{14} \: \mbox{W}/\mbox{cm}^2$ while varying the repeller voltage from 10 V to a maximum of 80 V. The peak intensity is changed by inserting pellicle beamsplitters, which have a negligible effect on pulse duration. The laser intensity is independently calculated prior to each data set by measuring the optical power using a thermal power meter, the pulse duration using an auto correlator and the focused spot size using a charged-coupled device (CCD) beam profiler. The error in the independent measurement of absolute peak intensity is $\sim 10$\%, which is comparable to the state of the art \cite{Alnaser-Cocke-peak-intensity-calibration}.

As seen from Eq. (\ref{hextract}), quantitative characterization of the dissociation efficiency $\mu$ is needed in order to isolate the electron yield due to hydrogen. We use emission spectroscopy of the discharge to obtain an accurate value for $\mu$. The relative intensities of two atomic H lines, Balmer-$\alpha$ and Balmer-$\beta$, and one molecular line, (2-2)Q1 of the H$_2$ Fulcher-$\alpha$ system, give the degree of dissociation \cite{Lavrov-Ropcke-emission-spectrum-dissoc-fraction-expt}. Typical values of the intensity ratios $I_{B_\alpha}/I_{(2-2)Q1}$ and $I_{B_\beta}/I_{(2-2)Q1}$ are 393 $\pm$ 39 and 74 $\pm$ 7 respectively. A measurement of the rotational series of the (2-2)Q Fulcher-$\alpha$ system must also be made in order to calculate the effective gas temperature inside the discharge \cite{Astashkevich-Ropcke-H2-plasma-temp-measurement}. The gas temperature inferred from those measurements is 480 $\pm$ 50 K. All intensities are measured using a 1 m double monochromator with 0.15 nm resolution with detection by a photomultiplier tube. The measured dissociation fraction of $\mu = 80 \pm 15$\% atomic H by number is comparable to the dissociation fraction measurements in the literature \cite{Ding-Weise-microwave-discharge-H-source, Paolini-Khahoo-hydrogen-dissociation-source}. The estimated error in the dissociation fraction measurement is inferred from errors in the relative intensities of the spectral lines.

Our experimental data are compared to theoretical predictions obtained by direct integration of the non-relativistic time-dependent Schr{\"o}dinger equation (TDSE). Two different TDSE calculations were implemented independently. Both used the velocity form of the interaction Hamiltonian $\hat{H}$ and expanded the wavefunction in spherical harmonics $\Psi = \sum_{\ell = 0}^{\ell_\mathrm{max}} R_\ell(r) Y_{\ell 0}(\theta)$, with radial functions $R_\ell(r)$ defined in a box of size $R_\mathrm{max}$. The first method used the Lanczos propagator \cite{Park-Light-lanczos-wavepacket-propagator}. This method represents the wavefunction as a linear combination of vectors $\Psi, \hat{H}\Psi, \ldots, \hat{H}^m\Psi$, where $m = 5$. Results converged well with values of $\ell_\mathrm{max} = 25$ a.u. and $R_\mathrm{max} = 2000$ a.u. and time-step $\Delta = 0.01$ a.u. The second approach was the `matrix iteration method' \cite{Nurhuda-Faisal-matrix-iteration-TDSE}. Its specific application to the present problem has been described recently \cite{GrumGrzhimailo-Bartschat-H-ionization}. Results converged for $\ell_\mathrm{max} = 30$ a.u., $R_\mathrm{max} = 3000$ a.u., and $\Delta = 0.04$ a.u. Comparison of the predictions from the two TDSE approaches resulted in excellent agreement and only one set of TDSE simulations is used in the following.

The theoretical simulations give momentum-resolved electron probability distributions for a fixed laser peak intensity. To simulate the experimental signal, we integrated the theoretical distributions over the detector acceptance function, the Gaussian laser intensity distribution, and the atomic density distribution in the interaction region. Careful modeling of the detector acceptance and the laser intensity distribution was critical to obtain good agreement between experiment and theory; for instance, approximating the laser intensity by a top-hat profile led to disagreement at the 50\% level. The detector acceptance function was found by simulating the electron trajectories in the electrostatic fields of the detection apparatus for a variety of trajectory initial conditions and electrode voltages. The acceptance angle was $\approx 7^\circ$ with slight variations as a function of repeller voltage $V_D$. The electron cutoff energy $E_D$ was found to follow the approximate formula $E_D \approx 0.8(V_D - 5)$, where $E_D$ has units of eV and $V_D$ has units of V.

To account for the Gaussian intensity variation of the laser beam, the atomic density is taken to be constant within an infinitely long cylinder oriented perpendicular to the laser beam direction and zero outside this cylinder. The Rayleigh range of the focused laser beam is much larger than the atomic beam width, which is again much greater than the focused spot size. The overall electron yield is therefore equal to the integral of the detection probability over the transverse Gaussian profile of the laser beam \cite{Alnaser-Cocke-peak-intensity-calibration}.

Figure~2 shows the detected ionisation signal from atomic H compared to the TDSE prediction. The atomic H electron yield is plotted as a function of electron cutoff energy $E_D$ for a number of laser intensities, using the approximate linear relationship between repeller voltage $V_D$ and cutoff energy $E_D$. The theoretical predictions are compared to experimental data by a fitting routine that takes each independently measured intensity measurement as an input. The full set of data points is simultaneously fitted using only two fit parameters: an overall scaling factor and an intensity scaling factor. The constant overall scale factor applied to the yield accounts for the absolute detector efficiency and the target density in our apparatus, which are not independently measured.  The intensity scaling factor gives a fit value for the absolute peak intensity, while leaving the relative intensities of the various data runs at their independently measured values.

\begin{figure}
\centerline{\includegraphics*[width=\columnwidth]{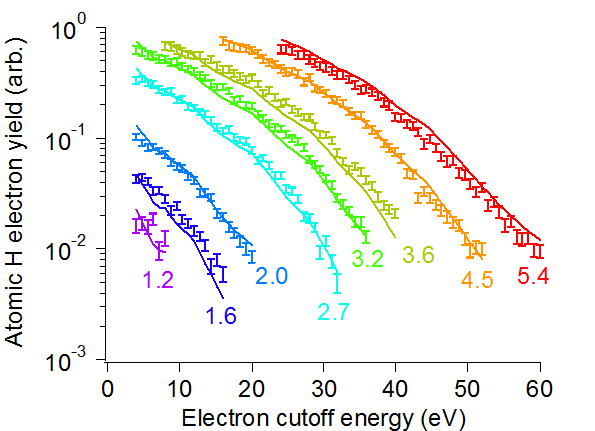}}
\caption{Experimental data (dots) versus TDSE predictions (solid line). The laser intensity range is 1.2 to 5.4 $\times 10^{14} \:\mbox{W}/\mbox{cm}^2$ and is indicated below the corresponding data run. The difference between two TDSE predictions was much smaller than our experimental error.}
\label{data}
\end{figure}

As shown in Fig. 2, we obtain agreement at the 10\% level between the experimental data and the {\it ab initio} TDSE prediction over a wide range of electron energies and laser intensities. Our analysis of experimental uncertainty includes contributions from the dissociation efficiency measurement, the shot noise, and an Allan deviation analysis of the electron yield. The uncertainty in determining $\mu$ results in a contribution of $\sim 5$\%, the shot noise is 4 - 10\% depending on yield, and an Allan deviation analysis shows that an additional $\sim 5$\% change in electron yield can be expected over the duration of data acquisition. The data run with an intensity of $3.2 \times 10^{14} \:\mbox{W}/\mbox{cm}^2$ was a slight outlier and a correction factor was found by fitting the run independently. The measured intensity was therefore adjusted by 6\%, an amount consistent with experimental error in the independent measurement of intensity. The excellent agreement between data and predictions over a wide range of parameters provides strong evidence of our ability to model the experimental conditions accurately.

The absolute intensity scale determined by the TDSE fit is 0.94 $\pm$ 0.01 times the independently measured absolute intensity scale. Since this scale factor is consistent with 1 to within our absolute intensity calibration error of 10\%, we see that the TDSE fit is capable of extracting the true peak intensity from the data at this level of uncertainty. The small error in the intensity scaling for the TDSE fit offers the possibility of laser intensity calibration at the 1\% level in future experiments.

We have presented experimental results on few-cycle, strong-field interaction with atomic H, the classic test system for atomic physics. The excellent quantitative agreement with the TDSE model indicates the accuracy of our model for our experimental conditions. Similar experiments can be accurately calibrated using atomic hydrogen and subsequently used with other atomic and molecular species. Future measurements of fully resolved electron momentum distributions and HHG spectra for atomic H will provide benchmarks for strong-field physics.

We thank Dan Kleppner for assistance with the atomic hydrogen beam source. This work was supported by the US Air Force Office of Scientific Research under grant FA2386-09-1-4015 and the Australian Research Council under grants DP0878560 and CE0561787.

\end{document}